\documentclass[aps,prb,twocolumn]{revtex4}
\usepackage{epsfig}
\usepackage{graphicx}
\begin{document} 
\title{Finite $U$-induced competing interactions, frustration, and quantum phase transition \\
in a triangular-lattice antiferromagnet}
\author{Avinash Singh}
\email{avinas@iitk.ac.in} 
\affiliation{Department of Physics, Indian Institute of Technology Kanpur - 208016}
\begin{abstract}
The $120^0$ ordered antiferromagnetic state of the Hubbard model on a triangular lattice 
presents an interesting case of $U$-controlled competing interactions and frustration.
The spin stiffness is found to vanish at $U^* _{\rm stiff} \approx 6$
and the spin-wave energy $\omega_M$ at ${\bf q}_M=(2\pi/3,0)$ etc. 
is found to vanish at $U_M ^* \approx 6.8$
due to competing spin couplings generated at finite $U$. 
The loss of magnetic order due to the magnetic instability at ${\bf q}_M$ 
yields a first-order quantum phase transition in the insulating state at $U=U_{\rm M}^*$. 
Implications of the quantum spin disordered insulator to the spin-liquid state and
Mott transition in the organic systems $\rm \kappa -(BEDT-TTF)_2 X$ are discussed.
Effects of hole and electron doping on magnetic ordering and spin stiffness 
are also examined.
\end{abstract}
\pacs{75.50.Pp,75.30.Ds,75.30.Gw}  
\maketitle
\section{Introduction}

Recent $^1$H NMR and static susceptibility measurements on the 
{\em nearly isotropic}, triangular-lattice antiferromagnet 
$\rm \kappa -(BEDT-TTF)_2 Cu_2  (CN)_3$ 
have shown no indication of long-range magnetic ordering down to 32 mK, 
well below the estimated exchange constant $J \sim 250$ K,
suggesting the realization of a quantum spin-liquid state.\cite{kanoda}
No signature of antiferromagnetic (AF) transition was seen 
in earlier EPR measurements as well.\cite{epr}

On the other hand, a non-colinear $120^0$ ordered ground state is the accepted consensus 
for the $S=1/2$ quantum Heisenberg antiferromagnet (QHAF) 
on an isotropic triangular lattice.\cite{order1,order2}
Recent quantum Monte Carlo calculations yield a quantum reduction of $59\%$ 
to the magnetic order from its classical value,\cite{qmc}
quite close to the spin-wave theory result of $52\%$ 
in the first-order $1/S$ expansion,\cite{loop1,loop2}
which is somewhat greater than the $40\%$ reduction for the unfrustrated square-lattice.

In this paper we show that the absence of long-range magnetic order in the 
nearly isotropic organic antiferromagnet $\rm \kappa -(BEDT-TTF)_2 Cu_2  (CN)_3$ 
can be understood in terms of strongly enhanced quantum spin fluctuations
due to finite $U$-induced competing interactions and frustration.
Indeed, for the half-filled Hubbard model on an isotropic triangular lattice,
we find that in the insulating ordered state the spin stiffness vanishes at 
$U^* _{\rm stiff} \approx 6$.
The corresponding divergence in the spin-fluctuation correction implies
loss of long-range magnetic order at $U^* _{\rm order}$ which is 
somewhat higher than $U^* _{\rm stiff}$,
yielding a finite-$U$ magnetic quantum phase transition in the insulating state. 
Even deeper in the insulating state, we find an instability of the $120^0$ ordered 
state at $U_M ^* = 6.8$ against out-of-plane spin fluctuations of 
wave vector ${\bf q}_M=(2\pi/3,0)$ etc., implying instability towards a F-AF state.
Since spin fluctuations remain finite as 
$U \rightarrow U_M ^*$, the magnetic instability implies a first-order
quantum phase transition, thus pre-empting the vanishing spin-stiffness instability.

The realization in the triangular lattice of a non-magnetic insulator state at 
intermediate $U$, in which magnetic ordering is suppressed by strong quantum spin
fluctuations, is interesting as it allows, with decreasing $U$,
for a Mott-type metal-insulator transition not accompanied by any magnetic symmetry
breaking. In this context, another quantum correction which assumes significance
is that for the AF band gap. Indeed, two different scenarios emerge depending on
the relative magnitudes of  $U^* _{\rm order}$ and $U^* _{\rm gap}$,
where the $120^0$ AF order and AF band gap vanish, respectively.
For $U^* _{\rm order} > U^* _{\rm gap}$,
a non-magnetic insulator state lies between the 
paramagnetic metal (PM) and the AF insulator (AFI), 
whereas for $U^*_{\rm gap} > U^*_{\rm order}$, 
the magnetic transition is pre-empted, and there is a (nearly)
first-order transition from
AFI to PM at $U^* _{\rm gap}$ when the two AF bands start overlapping,
possibly with an intervening antiferromagnetic metallic (AFM) phase
in a narrow $U$ range.
Quantum corrections to quasiparticle dispersion and band gap due to motion of an added
hole (electron) in the AF background, resulting in strong incoherence due to 
scrambling of the spin ordering, has been recently studied in detail for the 
$t-t'$-Hubbard model on a square lattice.\cite{self}
 
The divergent quantum spin fluctuations due to vanishing spin stiffness 
and the magnetic instability towards a F-AF state may actually
be a precursor to an exotic quantum spin-disordered state,
as for the frustrated square-lattice antiferromagnet.
Indeed, for the spin-half $J-J'$ Heisenberg model on a square lattice,
or equivalently the strong-coupling $t-t'$-Hubbard model,
where the frustrating NNN coupling $J'=4t'^2/U$ leads to vanishing spin stiffness
and a $(0,\pi)$ instability towards a F-AF state at $J'/J=t'^2/t^2=0.5$,\cite{phase} 
series-expansion studies\cite{rrpsingh,kotov} 
of the ground-state energy indicate a continuous transition at $J'/J \sim 0.4$ 
from the AF state which breaks spin-rotation symmetry
to a columnar dimer (valence-bond-solid) state 
which breaks lattice translation symmetry, 
although it is controversial.\cite{sorella}
The location of the transition at $J'/J \sim 0.4$
is surprisingly very close to where the AF order 
also vanishes.\cite{phase}
Recently there has been strong interest in the critical theory 
of continuous quantum phase transitions between two phases with different broken symmetry, 
which requires going beyond the Landau-Ginzburg-Wilson paradigm.\cite{ashvin}

The suppression of magnetic ordering due to enhanced spin fluctuations is therefore
also relevant in the layered system $\rm \kappa -(BEDT-TTF)_2 Cu[N(CN)_2]Cl$, 
which exhibits a genuine Mott transition 
not accompanied by any symmetry breaking.\cite{kagawa}
Generally, the organic systems $\rm \kappa -(BEDT-TTF)_2 X$, where $X$ denotes inorganic
monovalent anion, have emerged as a new class of correlated electron systems exhibiting
antiferromagnetism, superconductivity, and metal-insulator transition.\cite{review1,review2}
Recent discovery of superconductivity in $\rm Na_x Co O_2 . y H_2 O$ \cite{watersup}
and the observation of low-temperataure insulating phases 
in some $\sqrt{3}$-adlayer structures such as K on Si[111],\cite{weitering} 
have also renewed interest in correlated electron system on triangular lattices.
As a recent example of quasi-two-dimensional antiferromagnetism on a triangular lattice
exhibiting the $120^0$ spin ordering,  $\rm Rb Fe (Mo O_4)_2 $ has been studied 
using elastic neutron scattering,\cite{neutron}
and magnetic resonance and magnetization experiments.\cite{resonance}

The Hubbard model on a triangular lattice has been studied recently using a variety of tools. 
The non-magnetic insulating state near the Mott transition has been studied using the 
path integral renormalization group method,\cite{pirg} 
in which the HF results are systematically improved to reach the true ground state 
by taking account of quantum fluctuations.
Results show a generic emergence of a non-magnetic insulating state sandwiched by a
Mott metal-insulator transition and an AF transition. 
The zero-temperature phase diagram has been studied using the slave boson technique and
the exact diagonalization.\cite{slavebos}
The mean-field SB approach yields a rich phase diagram qualitatively resembling
the HF results.\cite{hf1,hf2}
One-electron density of states has been examined 
using the quantum Monte Carlo method,\cite{qmc_green}
showing a pseudogap development for intermediate $U$,
accompanied by two peaks in the spin structure factor, 
signaling the formation of a spiral spin density wave (SDW).
A weak-coupling RG analysis applied to the anisotropic triangular lattice shows 
that frustration suppresses the antiferromagnetic instability in favour of a superconducting
instability.\cite{weakcoup}
A magnetic field induced exotic spin-triplet superconductivity has been proposed
having strong ferromagnetic fluctuations.\cite{triplet}
Ground-state spin structure of Cr and Mn monolayers on Cu[111], proposed as 
ideal candidates for physical realization of frustrated 2D {\em itinerant} antiferromagnets,
has been investigated by performing {\em ab initio} calculations based on 
the density-functional theory in the local spin-density approximation.\cite{crmono}

A spin-liquid type non-magnetic insulating (NMI) state sandwiched between a weak-coupling
PM state and a strong-coupling AFI state has also been obtained for
the $t-t'$-Hubbard model on a square lattice and an anisotropic
triangular lattice using the path integral
renormalization group method.\cite{pirg2,pirg3}
The NMI state has been recently suggested to be a new type of
degenerate quantum spin phase having gapless and dispersionless spin
excitations.\cite{pirg3} At the same time,
this result of an intervening NMI state is in contradiction to the earlier finding
of an intermediate metallic antiferromagnetic state (AFM).\cite{duffy}
We briefly compare the two types of lattices to highlight the
common and distinguishing features of the AF state.

Frustration suppresses the spin stiffness in both cases,
rendering the AF state more susceptible to loss of magnetic order due to
quantum fluctuations.
While frustration is explicitly controlled by $t'$ in the square lattice
and can be tuned to zero, the triangular lattice provides a subtle case of
$U$-controlled intrinsic frustration where the stiffness can be tuned to zero
at finite $U$. 
Another common consequence of frustration is the AF band broadening,
which is due to same-sublattice hopping $t'$ in the square lattice
and due to the spiral ordering in the triangular lattice.
The consequent reduction of the AF band gap
also enhances the fluctuation correction due to interband transfer of
spectral weight.
Furthermore, the suppression of the perfect-nesting instability in both cases
has the result that magnetic ordering sets in only above a critical
interaction strength.
Whether the onset of magnetic ordering and opening of AF band gap occur
at different $U$ values,
yielding an intervening AFM phase for the triangualr lattice,
is a pertinent question.
At the HF level, an AFM phase is indeed stabilized,
although in a narrow $U$ range $4.7 < U < 5.1$,
indicating that the magnetic transition is nearly first order;
whereas for the square lattice with $t'/t \lesssim 0.5$,
the magnetic order collapses when the two AF bands start overlapping,
yielding a strictly first-order magnetic transition,
the AFM phase is again stabilized for the cubic lattice.\cite{duffy,hofstetter,phase}
This shows a subtle dependence on the density of overlapping band states,
indicating that density of states renormalization due to quantum fluctuations
should have an important effect on stabilization of the AFM phase.

We consider the Hubbard model 
\begin{equation}
{\cal H} =  -t \sum_{i,\delta} a_{i,\sigma} ^\dagger a_{i+\delta,\sigma} 
+ U \sum_i a_{i\uparrow} ^\dagger a_{i\uparrow} a_{i\downarrow} ^\dagger a_{i\downarrow}
\end{equation}
with nearest-neighbour (NN) hopping on a triangular lattice,
and focus on the effect of finite $U$-induced competing interactions 
on quantum spin fluctuations in the $120^0$ ordered state.
The mean-field state is briefly reviewed in section II to introduce the notation.
We obtain the spin fluctuation propagator (section III)
and study the spin-wave excitations and spin stiffness in the full $U$ range (section IV).
We also examine the effects of hole and electron doping on magnetic ordering (section V).

\section{Mean-field state}
There are two alternative mean-field descriptions of the $120^0$ ordered AF state ---
i) a spiral-state representation, 
with an ordering wavevector ${\bf Q}=(2\pi/3,2\pi/\sqrt{3})$, 
and ii) a three-sublattice representation,
involving the local mean fields ${\bf \Delta}_\alpha$ on the three sublattices
$\alpha=\rm A,B,C$.
The energy eigenvalues and eigenvectors of the sublattice-basis $[6\times 6]$ Hamiltonian matrix 
can be conveniently obtained from those of the spiral-state $[2\times 2]$ Hamiltonian, 
as described below.  
\subsection{Spiral-state representation} 
With an order parameter $\Delta_{\bf Q}=-U\sum_{\bf k} \langle a_{{\bf k-Q}\downarrow}^\dagger a_{{\bf k}\uparrow}\rangle$, representing spin ordering in the 
$x-y$ plane, the Hubbard Hamiltonian reduces to 
\begin{equation}
H_{\rm HF}=\sum_{\bf k} (a_{{\bf k}\uparrow}^\dagger
 \; a_{{\bf k-Q}\downarrow}^\dagger )
\left [ \begin{array}{lr}\epsilon_{\bf k} & \Delta_{\bf Q} \\
\Delta_{\bf Q}^* & \epsilon_{\bf k-Q} \end{array} \right ]
\left ( \begin{array}{l} a_{{\bf k}\uparrow} \\ a_{{\bf k-Q}\downarrow} \end{array} \right )
\end{equation}
at the Hartree-Fock (HF) level,
where $\epsilon_{\bf k} = -2t[\cos k_x + 2\cos (k_x/2) \cos (k_y \sqrt{3}/2) ]$
is the triangular-lattice free-fermion energy.
Choosing real $\Delta_{\bf Q}$,
the spiral-state quasiparticle energies $E_{\bf k}^\pm$ and amplitudes $(u_{\bf k} \; v_{\bf k})$ 
are given by
\begin{eqnarray}
E_{\bf k}^\pm &=& \eta_{\bf k} \pm \sqrt{\Delta_{\bf Q}^2 + \xi_{\bf k}^2} \\
u_{\bf k}^2 &=& \frac{1}{2}\left (1 \pm \frac{\xi_{\bf k}}{\sqrt{\Delta_{\bf Q}^2 + \xi_{\bf k}^2}} \right ) \nonumber \\
v_{\bf k}^2 &=& \frac{1}{2}\left (1 \mp \frac{\xi_{\bf k}}{\sqrt{\Delta_{\bf Q}^2 + \xi_{\bf k}^2}} \right ) 
\end{eqnarray}
for the upper (+) and lower ($-$) AF bands,
where $\eta_{\bf k} \equiv (\epsilon_{\bf k} + \epsilon_{\bf k-Q})/2$
and $\xi_{\bf k} \equiv (\epsilon_{\bf k} - \epsilon_{\bf k-Q})/2$.
Self-consistency requires that 
$\Delta_{\bf Q}= -U \sum_{\bf k} v_{\bf k}^* u_{\bf k} $,
which yields the condition
\begin{equation}
\frac{1}{U}=\sum_{\bf k} \frac{1}{2\sqrt{\Delta_{\bf Q}^2 + \xi_{\bf k}^2}}
[\theta(E_{\rm F} - E_{\rm k}^- ) - \theta(E_{\rm F} - E_{\rm k}^+ ) ]
\end{equation}
in which lower and upper band states contribute with opposite sign.

In terms of the fermion spinor
$\Psi_i = (a_{i\uparrow} \; a_{i\downarrow})$,
the spin expectation values $\langle S^\mu _i \rangle = \frac{1}{2}
\langle \Psi_i ^\dagger \sigma^\mu \Psi_i \rangle$
in the spiral state yield
\begin{eqnarray}
\langle S^x _i \rangle = \frac{1}{2} m_{\bf Q} \cos {\bf Q.r_i} \nonumber \\
\langle S^y _i \rangle = \frac{1}{2} m_{\bf Q} \sin {\bf Q.r_i} 
\end{eqnarray}
at lattice site $i$,
where the spiral-state magnetization 
$m_{\bf Q}= 2 \sum_{\bf k} \langle a_{{\bf k-Q}\downarrow}^\dagger a_{{\bf k}\uparrow}\rangle$.
For ${\bf Q}=(2\pi/3,2\pi/\sqrt{3})$ the spiral twisting of spins generates the $120^0$
ordered AF state on the triangular lattice.

\subsection{Three-sublattice representation} 
While the spiral-state description applies only to Bravais lattices,
the sublattice-basis description applies to Kagom\'{e} type non-Bravais lattices as well.
In the Hartree-Fock approximation, the Hamiltonian reduces to
\begin{equation}
{\cal H}_{\rm HF} = \sum_i 
\Psi_i^\dagger [-{\mbox{\boldmath $\sigma$}}.{\bf \Delta}_i]
\Psi_i  - t \sum_{i,\delta} 
\Psi_i^\dagger {\bf 1} \Psi_{i+\delta} \; ,
\end{equation}
where  the local mean field ${\bf \Delta}_i = \frac{1}{2}
U\langle \Psi_i ^\dagger {\mbox{\boldmath $\sigma$}} \Psi_i \rangle $.
In general,
the $120^0$ AF state is characterized by an ordering plane (normal $\hat{n}_1$) 
and a planar direction ($\hat{n}_2$) with reference to which  ${\bf \Delta}_i$ 
make angles $\theta_\alpha=0^0$, 120$^0$, and 240$^0$
on the three sublattices $\alpha=\rm A,B,C$.
A convenient choice is $\hat{n}_1=\hat{z}$ 
(spin-ordering in the $x-y$ plane) and $\hat{n}_2=\hat{x}$, 
so that ${\bf \Delta}_i$ on the three sublattices are given by
\begin{equation}
{\bf \Delta}_\alpha = \Delta \hat{\alpha} \;\;\;\;\; (\hat{\alpha}=\hat{a},\hat{b},\hat{c})
\end{equation}
in terms of the three lattice unit vectors 
\begin{equation}
\hat{a} = \hat{x}, \;\;\;\; 
\hat{b} = -\frac{1}{2}\hat{x} + \frac{\sqrt{3}}{2} \hat{y}, 
\;\;\;\;\;
\hat{c} = -\frac{1}{2}\hat{x} - \frac{\sqrt{3}}{2} \hat{y} .
\end{equation}
In general, in terms of the two mutually perpendicular directions
$\hat{n}_1$ and $\hat{n}_2$, the spin orientations in Eq. (8) are given by
\begin{equation}
\hat{\alpha} = \cos\theta_\alpha \; \hat{n}_2
+ \sin\theta_\alpha (\hat{n}_1 \times \hat{n}_2 ).
\end{equation}

As ${\bf \Delta}_i$ is identical on sites of the same sublattice,
Fourier transformation within the sublattice basis yields
\begin{equation}
{\cal H}_{\rm HF} = \sum_{\bf k} \Psi_{\bf k} ^\dagger
\left [ \begin{array}{ccc}
-{\bf \sigma}.{\bf \Delta}_{\rm A} & \delta_{\bf k} & 
\delta_{\bf k}^*  \\
 & & \\
\delta_{\bf k}^* & -{\bf \sigma}.{\bf \Delta}_{\rm B} & 
\delta_{\bf k} \\
 & & \\
\delta_{\bf k}  & \delta_{\bf k}^* & 
-{\bf \sigma}.{\bf \Delta}_{\rm C} 
\end{array}
\right ]\Psi_{\bf k} .
\end{equation}
Here $\Psi_{\bf k} \equiv 
(a_{{\bf k}\uparrow}\;a_{{\bf k}\downarrow}
\; b_{{\bf k}\uparrow}\;b_{{\bf k}\downarrow}
\; c_{{\bf k}\uparrow}\;c_{{\bf k}\downarrow} )$,
where $a_{\bf k},b_{\bf k},c_{\bf k}$ are fermion operators defined on the three 
sublattices A, B, C. Wavevector $\bf k$ lies within the Magnetic Brillouin Zone (MBZ),
corresponding to the three inter-penetrating triangular sublattices (lattice parameter $\sqrt{3}a$). 
The NN hopping term 
\begin{equation}
\delta_{\bf k} = -t \sum_{\hat{\delta}=\hat{a},\hat{b},\hat{c}}
 e^{i{\bf k}.\hat{\delta} }
= -t [e^{i k_x} + 2e^{-ik_x/2} \cos (\sqrt{3} k_y/2) ]
\end{equation} 
mixes AB, BC, and CA sublattices, 
which are connected by the three lattice unit vectors. 

The six eigenvalues $E_{{\bf k},l}^\pm$ and eigenvectors $|{\bf k}^\pm,l\rangle $ of the 
$[6\times 6]$ Hamiltonian matrix in Eq. (11),
corresponding to upper ($+$) and lower ($-$) AF bands,
follow from Eqs. (3) and (4) for the spiral-state Hamiltonian. 
Here $l=1,2,3$ refer to the three branches corresponding to momentum values 
$\bf k$, $\bf k+Q$, $\bf k-Q$, respectively, $\bf k$ being restricted to the MBZ. 
The amplitude $|{\bf k},l \rangle_\alpha$ involves not only the spin orientation $\phi_\alpha$ corresponding to sublattice $\alpha$, but also a relative phase angle $\delta_{\bf Q} ^{l,\alpha}$ associated with the spiral twisting, and is given by
\begin{equation}
|{\bf k},l \rangle_\alpha = \left ( \begin{array}{l} 
u_{{\bf k},l} \; e^{-i\phi_\alpha} \\
v_{{\bf k},l} \;  \end{array} \right )
e^{i\delta_{\bf Q} ^{l,\alpha}}
\end{equation}
where the planar spin orientations $\phi_\alpha = 0^0,\; 120^0,\; 240^0$ for the three 
sublattices. The spiral phase angle 
\begin{equation}
\delta_{\bf Q} ^{l,\alpha} = 0, \; \pm {\bf Q}.{\bf R}_{\alpha \rm A}
\end{equation}
(relative to A) for $l=1,2,3$, where ${\bf R}_{\alpha \rm A}$ is the primitive lattice vector connecting sublattices $\alpha$ and A.
In view of above structure of state $|{\bf k}\rangle$, 
the self-consistency condition 
$m^\mu _\alpha= 2\Delta^\mu _\alpha /U = \sum_{{\bf k},l} 
\langle {\bf k},l | \sigma^\mu | {\bf k},l \rangle_\alpha $ 
retains the same form as in Eq. (5). 

\section{Spin fluctuations}
Associated with the spontaneous symmetry breaking of the 
continuous spin-rotation symmetry of the Hubbard model,
there exist gapless transverse fluctuation modes or Goldstone modes,
involving fluctuations locally normal to the symmetry-breaking directions.
These low-energy collective excitations,
studied earlier in the context of doped cuprates,\cite{collective}
play an important role in determining several important physical properties 
in ordered systems, such as stability of the mean-field ordered state,
quantum corrections to the order parameter and ground-state energy,
temperature dependence of the order parameter in three dimensions,
and renormalization of electron (hole) spectral function and AF band gap 
due to fermion-magnon scattering of mobile carriers in the magnetic background.

We consider the time-ordered spin fluctuation propagator, the Fourier transform of which is given in the sublattice basis as 
\begin{eqnarray}
[\chi ({\bf q},\omega)]^{\mu\nu} _{\alpha \beta} &=&
i \int dt  \; e^{i\omega (t-t')} \sum_j e^{i{\bf q}.({\bf r}_i - {\bf r}_j)} \nonumber \\
&\times & \langle \Psi_{\rm G} |{\rm T}[
S_i ^\mu (t) S_j ^\nu (t')]\Psi_{\rm G} \rangle
\end{eqnarray}
where $\alpha,\beta= \rm A,B,C$ are the sublattice indices 
and $\mu,\nu=x,y,z$ are the spin directions.
To leading order in an inverse-degeneracy expansion,\cite{quantum} 
equivalent to summing over all bubble diagrams, 
\begin{equation}
[\chi ({\bf q},\omega)]  = \frac{\frac{1}{2}\;[\chi^0 ({\bf q},\omega)]}
{{\bf 1} - U[\chi^0 ({\bf q},\omega)]}
\end{equation}
where the bare particle-hole propagator $[\chi^0 ({\bf q},\omega)]$ is 
obtained by integrating out the fermions in the broken-symmetry state.
For the half-filled insulating state, the added hole (particle) states lie in the 
lower (upper) Hubbard band, and we have
\begin{equation}
[\chi^0({\bf q},\omega)]^{\mu\nu} _{\alpha \beta} = \frac{1}{2}
\sum_{{\bf k}lm}
\frac{\langle \sigma^\mu \rangle_\alpha ^{-+}
\langle \sigma^\nu \rangle_\beta^{-+*}}
{E_{\bf k-q}^+ - E_{\bf k}^- + \omega} 
+ 
\frac{\langle \sigma^\mu \rangle_\alpha ^{+-}
\langle \sigma^\nu \rangle_\beta ^{+-*}}
{E_{\bf k}^+ - E_{\bf k-q}^- - \omega}  \; ,
\end{equation}
where $\langle \sigma^\mu \rangle_\alpha ^{-+}$ denotes 
the particle-hole spin matrix element on the $\alpha$ sublattice
\begin{equation}
\langle \sigma^\mu \rangle_\alpha ^{-+} \equiv
\langle {\bf k-q}^+ ,m| \sigma^\mu |{\bf k}^- ,l\rangle_\alpha.
\end{equation}

The bare particle-hole propagator 
$[\chi^0({\bf q},\omega)]^{\mu\nu} _{\alpha,\beta}$
is a $[9 \times 9]$ Hermitian matrix in the three sublattice ($\alpha$=A,B,C) 
and three spin ($\mu=x,y,z$) basis, the (real) eigenvalues $\lambda_{\bf q}(\omega)$ 
and eigenvectors $|\phi_{\bf q}(\omega)\rangle $ of which 
contain information regarding the massless transverse fluctuations (spin waves) 
as well as the massive longitudinal fluctuations along the local ordering direction.
Also included are the particle-hole (Stoner) excitations across the band gap,
given by the poles of $[\chi^0({\bf q},\omega)]^{\mu\nu} _{\alpha,\beta}$.
In the following, we focus on the spin-wave energies $\omega_{\bf q}$, 
which are obtained from the poles $1-U\lambda_{\bf q}(\omega_{\bf q}) = 0$ 
of Eq. (16).

Evaluating and diagonalizing the $[\chi^0({\bf q},\omega)]$ matrix for ${\bf q},\omega=0$,
we obtain three Goldstone modes $U\lambda_n = 1$, 
as expected for the non-colinear $120^0$ AF state,
corresponding to rigid spin rotations about the $x,y,z$ axes.
Rotation around the $z$ axis yields an in-plane mode involving $S^x$ and $S^y$ fluctuations,
while those around the $x$ and $y$ axes yield two out-of-plane modes involving only $S^z$ fluctuation. 
The corresponding eigenvectors, giving transverse fluctuation amplitudes on the three sublattices, are 
\begin{equation}
\left (
\begin{array}{r}
\hat{y} \\  -\frac{\sqrt{3}}{2} \hat{x} -\frac{1}{2} \hat{y} \\
\frac{\sqrt{3}}{2} \hat{x} -\frac{1}{2} \hat{y} \end{array}\right ) \;\;\;
\left (
\begin{array}{r}
0 \\  \hat{z} \\ -\hat{z} \end{array}\right ) \;\;{\rm and}
\left (
\begin{array}{r}
1 \\  \frac{1}{2}\hat{z} \\ -\frac{1}{2}\hat{z} \end{array}\right ) 
\end{equation}

For small ${\bf q},\omega$, the Goldstone mode eigenvalue $\lambda_{\bf q}(\omega)$
has the following typical form for the AF state:
\begin{equation}
\lambda_{\bf q}(\omega)=\frac{1}{U} -{\cal A} q^2 + {\cal B} \omega^2 \; ,
\end{equation}
where the coefficient ${\cal A}$ of the $q^2$ term is proportional to the
spin stiffness $\rho$. The pole equation
$1-U\lambda=0$ yields the spin-wave energy $\omega_{\bf q}=cq$,
where the spin-wave velocity $c=\sqrt{{\cal A}/{\cal B}} \propto \sqrt{\rho}$ 
is related to the spin stiffness. 

\begin{figure}
\vspace*{-70mm}
\hspace*{-38mm}
\epsfig{figure=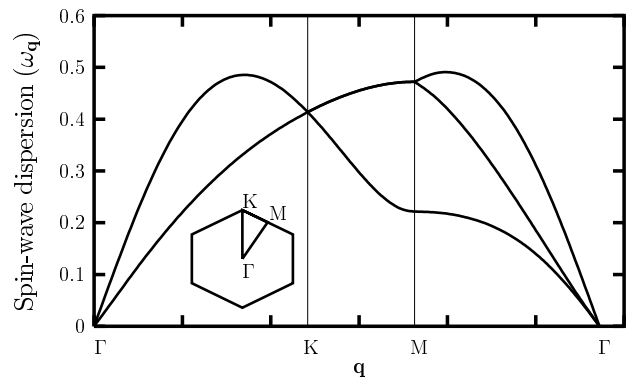,width=140mm}
\vspace{-78mm}
\caption{The spin-wave dispersion along different symmetry directions
of the MBZ (inset), for $\Delta=4$ ($U=8.8$).}
\end{figure}
\begin{figure}
\vspace*{-134mm}
\hspace*{-50mm}
\epsfig{figure=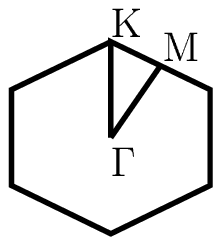,width=140mm}
\vspace{-78mm}
\end{figure}

\section{Spin-wave spectrum}
The full spin-wave dispersion along different symmetry directions in the 
Magnetic Brillouin Zone is shown in Fig. 1 for $U=8.8$,
corresponding to a self-consistent mean-field $\Delta=4$.
Along the $\Gamma$-K direction, the softer mode is doubly degenerate,
and the dispersion continues along the K-M direction after crossing at 
the MBZ vertex K, where all three modes become degenerate. 
The degeneracies along $\Gamma$-K and K-M directions in
${\bf k}$ space are simply related to the two equivalent
120$^0$ orderings (clockwise and anticlockwise rotation of spins)
along the corresponding orthogonal lattice directions
$\hat{x}$ and $\frac{1}{2}\hat{x} + \frac{\sqrt{3}}{2}\hat{y}$.
The degeneracy is resolved along the M-$\Gamma$ direction,
but two modes again become degenerate near $\Gamma$,
and these represent the two out-of-plane $(S_z)$ fluctuation modes.
Near the $\Gamma$ point ($q<<1$),
the $z$ sector does not mix with the $x-y$ sector in $[\chi^0({\bf q},\omega)]$,
and the sublattice symmetry in $z$ sector yields the two degenerate
$S_z$ fluctuation modes given in Eq. (19).
Behaviour of $\omega_{\bf q}$ near the special MBZ points
$\Gamma$, K, and M is further discussed below. 

We next consider the spin-wave energy at the MBZ vertices K 
with ${\bf q}_K=(0,4\pi/3\sqrt{3})$ etc., 
where all three spin-wave modes are degenerate,
allowing for a convenient comparison with the strong-coupling result.
In the strong-coupling limit $(U/t \rightarrow \infty)$, 
the spin-wave energies are given by:
\begin{equation}
\omega_{\bf q}=3JS[(1-\gamma_{\bf q})(1+2\gamma_{\bf q})]^{1/2}
\end{equation}
where $\gamma_{\bf q}=\frac{1}{3}[\cos q_x + 2 \cos(q_x/2) \cos (q_y \sqrt{3}/2)]$,
and the three modes correspond to momenta ${\bf q}$, ${\bf q+Q}$, and ${\bf q-Q}$.
As $\gamma_{\bf q}=0$ for all three modes for ${\bf q}_K=(0,4\pi/3\sqrt{3})$, 
the spin-wave modes are three-fold degenerate, with $\omega_K=3JS=6t^2/U$ for $S=1/2$.
Figure 2 shows the variation of $\omega_K$ with $U$,
along with the strong-coupling result for comparison. 
With decreasing $U$, the spin-wave energy turns over,
and the decreasing band gap effectively squeezes the spin-wave spectrum.
Indeed, both $\omega_K$ and the band gap vanish together
at $U\approx 5$ ($\Delta=2$).
This is a typical weak-coupling dynamical effect 
for an itinerant antiferromagnet.\cite{weak}
The divergence in the eigenvalue $\lambda_{\bf q}(\omega)$ of the  
$[\chi^0({\bf q},\omega)]$ matrix in Eq. (17), when spin-wave energy approaches 
the band gap, not only limits the spin-wave spectrum to within the band gap,
but also strongly suppresses the spin-wave amplitude due to 
wave-function renormalization.

\begin{figure}
\vspace*{-70mm}
\hspace*{-38mm}
\epsfig{figure=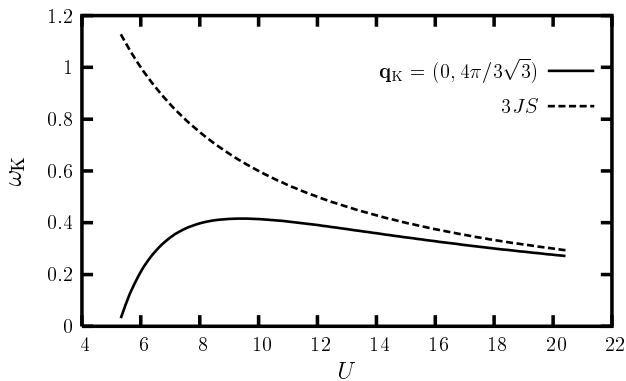,width=140mm}
\vspace{-78mm}
\caption{The spin-wave energy $\omega_{\rm K}$ at the Magnetic Brillouin Zone vertices K,
along with the strong-coupling result $3JS$.}
\end{figure}

\begin{figure}
\vspace*{-70mm}
\hspace*{-38mm}
\epsfig{figure=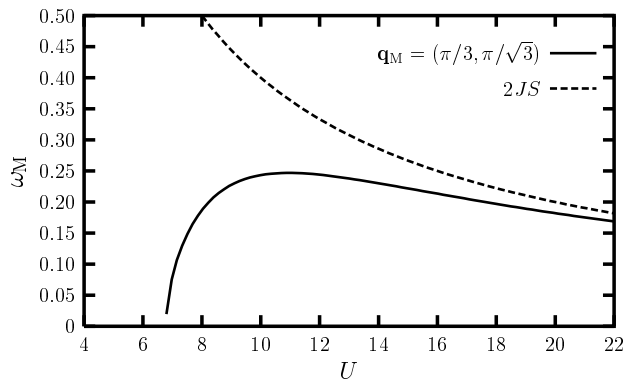,width=140mm}
\vspace{-78mm}
\caption{The spin-wave energy $\omega_{\rm M}$ at the mid-points (M) 
of the MBZ edges, along with the strong-coupling result $2JS$.}
\end{figure}

While the spin-wave energy $\omega_{\rm K}$ at the MBZ vertices K
vanishes along with the gap at $U\approx 5$,
the spin-wave energy $\omega_{\rm M}$ at the mid-points (M) of the MBZ edges 
shows evidence of competing interactions.
Figure 3 shows the behaviour of $\omega_{\rm M}$ with $U$ for 
the soft, non-degenerate mode (see Fig. 1),
along with the strong-coupling result $2JS$.
With decreasing $U$, $\omega_{\rm M}$ falls rapidly and vanishes at $U_M ^* =6.8$.
For $U<U_M ^*$, the maximum eigenvalue of $[\chi^0({\bf q}_{\rm M})]$ exceeds $1/U$,
signalling an instability of the $120^0$ ordered state.
This instability is driven by out-of-plane fluctuations,
as the corresponding instability eigenvector $\phi_\alpha ^\mu$ has
non-vanishing amplitudes only for spin direction $\mu=z$, with identical magnitudes for
all three sublattices $\alpha=A,B,C$. 
Analysis of the fluctuation amplitudes in the instability eigenvector 
indicates spin twisting in the $+z$ and $-z$ directions 
along alternating spin chains in the three lattice directions, 
implying instability towards a F-AF state,
as in the square-lattice AF with NNN hopping $t'$. 

The spin-wave dispersion $\omega_{\bf q}$ along the K-M-K direction [Fig. 4]
shows a cross-over from a quadratic behaviour around M for $U>U_M ^*$ to
a linear behaviour as $U\rightarrow U_M ^* = 6.8$.
For $U < U_M^*$, there are
no spin-wave solutions near M as $\lambda_{\bf q} > 1/U$.
Despite the instability at $U=U_M^*$, 
the fluctuation remains finite as $U\rightarrow U_M^*$.
This is because the magnon dispersion $\omega_{\bf q}$ near M becomes 
linear in (small) momentum difference $\tilde{\bf q}\equiv {\bf q} - {\bf q}_M$, 
and the fluctuation contribution of these modes
($\int \tilde{q} d\tilde{q} /c\tilde{q} $) remains finite, 
just as for the Goldstone modes. 
A discontinuous magnetic transition at $U=U_M ^*$ is therefore clearly indicated.

\begin{figure}
\vspace*{-70mm}
\hspace*{-38mm}
\epsfig{figure=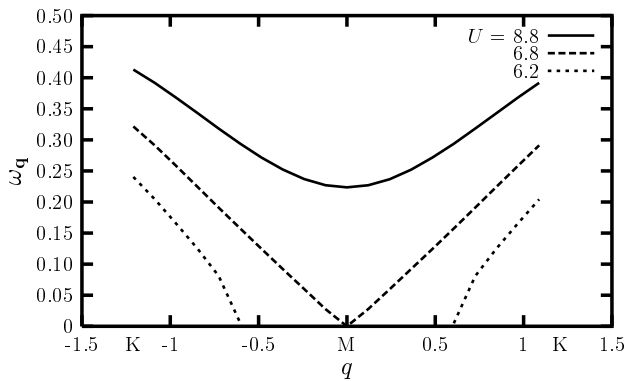,width=140mm}
\vspace{-78mm}
\caption{The spin-wave dispersion $\omega_{\bf q}$ along the K-M-K direction,
showing the cross-over from quadratic to linear behaviour around M
as $U\rightarrow U_M ^* = 6.8$.}
\end{figure}

\begin{figure}
\vspace*{-70mm}
\hspace*{-38mm}
\epsfig{figure=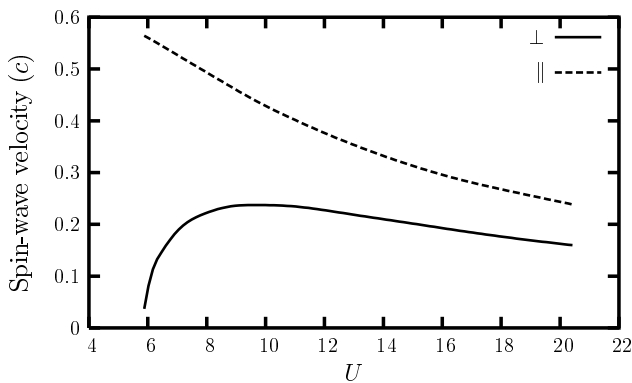,width=140mm}
\vspace{-78mm}
\caption{The spin-wave velocities for in-plane ($\parallel$) and out-of-plane $(\perp$)
fluctuation modes, showing that $c_\perp$ vanishes at $U \approx 6$. 
The ratio approaches $\sqrt{2}$ in the strong coupling limit.}
\end{figure}

Turning now to long-wavelength modes,
we consider the two spin-wave velocities $(c=\omega_{\bf q}/q)$
corresponding to in-plane $(c_\parallel)$ 
and out-of-plane $(c_\perp)$ fluctuations. 
For ${\bf q}=0$, the off-diagonal matrix elements of $[\chi^0 ({\bf q},\omega)]$, 
involving in-plane $(\mu=x,y)$ and out-of-plane $(\nu=z)$ spin indices, vanish identically, 
implying no mixing between the in-plane mode of the $x-y$ sector
and the two out-of-plane modes of the $z$ sector.
Mixing is negligible for $q << 1$ as well, 
and therefore small-$q$ spin-wave modes can also be identified 
in terms of in-plane and two out-of-plane fluctuations. 
In view of the exact degeneracy along the $\Gamma-K$ direction, 
it is convenient to consider ${\bf q} = (0,q_y)$, 
as the two spin-wave energies then readily yield the two spin-wave velocities,
as described below in the strong-coupling limit.

For $q_x=0$, the two modes ${\bf q} \pm {\bf Q}$ yield identical values:
$\gamma_{\bf q}=-\frac{1}{3}[1/2 + \cos(q_y \sqrt{3}/2)]$, and hence 
degenerate spin-wave energies from Eq. (21). For small $q_y$, one obtains 
$\omega_{\bf q}=3JS(\sqrt{3}/2)q_y$ and $3JS(\sqrt{3}/2\sqrt{2})q_y$
for the modes ${\bf q}$ and ${\bf q} \pm {\bf Q}$, respectively, 
yielding spin-wave velocities 
$c_\parallel=3JS(\sqrt{3}/2)$ and $c_\perp=3JS(\sqrt{3}/2\sqrt{2})$,
which are in the ratio $c_\parallel / c_\perp = \sqrt{2}$.

Figure 5 shows the two spin-wave velocities in the full $U$ range.
Both velocities decrease as $1/U$ for large $U$, 
and the ratio $c_\parallel / c_\perp$
asymptotically approaches $\sqrt{2}$ in the strong-coupling limit.
The intermediate-$U$ behaviour for the spin-wave velocity corresponding to 
out-of-plane fluctuations is most interesting, 
which exhibits a broad peak and falls rapidly with decreasing $U$, 
vanishing at a critical interaction strength $U^*_{\rm stiff}\approx 6$,
which is in the insulating regime.
For $U< U^*_{\rm stiff}$, the $120^0$ AF state is therefore 
unstable with respect to out-of-plane fluctuations. 

The vanishing spin-wave velocity and spin stiffness $(\rho_\perp \propto c_\perp ^2)$
implies that the first-order quantum correction 
to sublattice magnetization due to transverse spin fluctuations diverges. 
Therefore, the corrected sublattice
magnetization will vanish at a somewhat higher critical interaction strength 
$U^* _{\rm order} \gtrsim U^* _{\rm stiff}$,
where the quantum reduction due to fluctuations
exactly eliminates the mean-field order.
Hence there is a quantum phase transition in the triangular-lattice AF
at $U=U^* _{\rm order}$ from a $120^0$ ordered state
to a spin-disordered state.
As described below, this QPT is driven by  finite $U$-induced 
competing interactions and frustration. 

In the strong-coupling limit, 
the Hubbard model at half filling maps to the $S=1/2$ QHAF with NN interactions. 
However, for finite $U$, extended-range spin couplings
(of order $t^4/U^3$ and higher) are generated.
Within the RPA analysis, the spin couplings are approximately given by
$J_{ij} \approx U^2 [\chi^0 (\omega=0)]_{ij}$,
and have been obtained for the square lattice from a systematic $t/U$
expansion.\cite{coupling}
Now, for the unfrustrated AF, the O$(t^4/U^3)$ couplings  
between sites separated by two hoppings (NNN and NNNN) are ferromagnetic, 
which actually stabilize the AF state 
against the $t'$-induced frustration.\cite{phase}
However, for the 120$^0$ state on the triangular lattice, 
the extended-range ferromagnetic couplings are 
a source of additional frustration, leading to spin softening.

\begin{figure}
\vspace*{-70mm}
\hspace*{-38mm}
\epsfig{figure=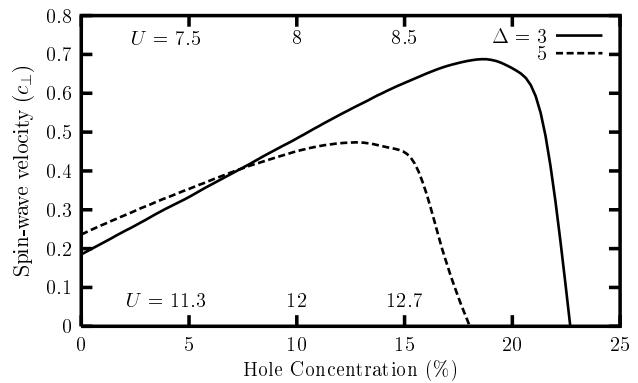,width=140mm}
\vspace{-78mm}
\caption{Enhancement in $c_\perp$ with hole doping,
due to a suppression of the intrinsic frustration in the triangular lattice.}
\end{figure}

\section{Hole and electron doping}
The two AF bands in the $120^0$ state are quite asymmetrical, 
with very different Fermi surfaces for hole and electron doping, 
suggesting quite different behaviour.
Indeed, we find that while hole doping stabilizes the AF state and actually 
increases the spin stiffness, any electron doping destroys AF ordering.

Stability of the square-lattice AF state for hole and electron doping was 
studied earlier within the $t-t'$-Hubbard model.\cite{doped}
For finite doping, the Fermi energy lies within a band,
and a key role is played by the {\em intraband} particle-hole processes 
in Eq. (17) for $[\chi^0({\bf q},\omega)]$,
which generate additional frustrating spin couplings and affect 
the stability of the AF state with respect to transverse spin fluctuations.
For positive $t'$, the AF state is destroyed for any electron doping, 
while a finite concentration is required for hole doping.

\begin{figure}
\vspace*{-70mm}
\hspace*{-38mm}
\epsfig{figure=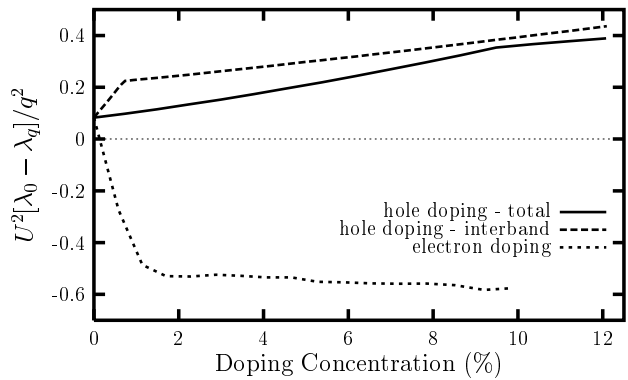,width=140mm}
\vspace{-78mm}
\caption{
Spin-stiffness contributions corresponding to
total (interband+intraband) and interband particle-hole processes,
for $\Delta=3$.
The spin stiffness increases with hole doping (solid) whereas it
becomes negative for electron doping (dotted).}
\end{figure}

We find a similar doping behaviour for the triangular lattice.
For hole doping in the (broadened) lower AF band
in a circular hole pocket around ${\bf k}=(0,0)$,
the spin-wave velocity $c_\perp$ initially {\em increases} (Fig. 6),
and then rapidly falls to zero with further hole doping,  
indicating destruction of the AF state. 
The initial enhancement in $c_\perp$ is due to a suppression of the intrinsic
frustration in the triangular lattice, as explained below.
In order to highlight the role of doping,
we have kept $\Delta$ rather than $U$ fixed, so that the band aspects
remain unchanged with doping. Over a small doping range $U$ remains nearly
same, as shown in the figure.

Figure 7 shows the spin-stiffness contributions corresponding to
total (interband $+$ intraband) and interband particle-hole processes in
$[\chi^0({\bf q},\omega=0)]$.
Here $\Delta=3$ and the $U$ values range between 7 and 8.
Interestingly, hole doping is seen to enhance the interband contribution,
reflecting an effective suppression of the intrinsic frustration by eliminating
the contribution of long-wavelength states in the lower AF band.
While the intraband contribution (total $-$ interband) does tend to
destabilize the AF state (negative stiffness),
the dominant interband contribution causes a net stabilization. 
A similar linear increase in the interband contribution with hole doping
was obtained in a detailed analytical study of the $t-t'$-Hubbard model.\cite{doped}

In contrast, for electron doping in the (narrow) upper AF band 
in elliptical electron pockets located symmetrically on the MBZ edges
at ${\bf k} = {\bf Q}/2 = (\pm\pi/3,\pm\pi/\sqrt{3})$ and $(\pm 2\pi/3,0)$,
the spin stiffness abruptly turns negative,
indicating instability of the $120^0$ AF state for any electron doping.
Furthermore,
the spin stiffness is seen to be quite independent of doping concentration.
Very similar results for the $t-t'$-Hubbard model were obtained
for electron doping in the elliptical electron pockets around
${\bf k} = (\pm\pi/2,\pm\pi/2)$.\cite{doped,affl}
The large relative magnitude of the intraband contribution is
characteristic of the highly elliptical Fermi surface which reduces
the energy denominator in the intraband particle-hole process.

\section{Conclusions}
In conclusion, 
spin-wave excitations were studied in the $120^0$ AF state 
of the Hubbard model on a triangular lattice
at half filling as well as for electron and hole doping. 
The triangular-lattice antiferromagnet presents a novel case
of $U$-controlled competing interactions and frustration, 
in contrast to the square-lattice case where frustration arises from the 
NNN coupling generated by the hopping term $t'$. 
The spin-wave energy $\omega_{\rm M}$ vanishes at $U_M ^* = 6.8$,
and the instability of the $120^0$ ordered state against out-of-plane fluctuations 
implies a first-order quantum phase transition in the insulating state 
which pre-empts the vanishing spin-stiffness instability involving
divergent quantum spin fluctuation and a continuous phase transition. 
The suppression of magnetic ordering due to enhanced quantum spin fluctuations
arising from the finite $U$-induced frustration
provides an explanation for why no long-range magnetic ordering is seen in
the nearly isotropic triangular-lattice antiferromagnet 
$\rm \kappa -(BEDT-TTF)_2 Cu_2  (CN)_3$. 
Furthermore, the realization of a non-magnetic insulating state at intermediate $U$,
which allows for a non-magnetic insulator - paramagnetic metal transition
when the AF band gap vanishes with decreasing $U$ and the two bands start
overlapping,
is relevant for the layered system $\rm \kappa -(BEDT-TTF)_2 Cu[N(CN)_2]Cl$, 
which exhibits a Mott-type metal-insulator transition 
not accompanied by any magnetic symmetry breaking.

A highly asymmetric doping behaviour was obtained for electron and hole doping,
when the {\em intraband} particle-hole processes
were incorporated in the spin-fluctuation propagator.
While hole doping was found to initially stabilize the AF state and actually 
increase the spin stiffness by suppressing the intrinsic frustration,
any electron doping was found to destroy the $120^0$ AF ordering. In fact,
the stabilization for hole doping and instability for electron doping,
with nearly concentration independent negative spin stiffness,
closely resembles the doping behaviour of the square-lattice AF
within the $t-t'$-Hubbard model for positive $t'$.

Finally, whether a non-magnetic insulator intervenes between the 
paramagnetic metal and the AF insulator depends on the relative strengths 
of the quantum corrections to magnetic order and AF band gap.
In case the AF band gap closes before the magnetic order is lost,
the magnetic transition is pre-empted, leaving a (nearly)
first-order AFI - PM transition,
when the AF bands start overlapping, possibly with an intervening
AFM phase in a narrow $U$ range.
In this context, quantum corrections to quasiparticle dispersion and band gap 
due to motion of an added hole (electron) in the $120^0$ ordered AF background 
are presently under investigation. 

\section*{Acknowledgement}
Helpful discussions with Z. Te\v{s}anovi\'{c} and V. Ashwin are gratefully acknowledged.

\end{document}